\documentclass{aa}
\usepackage{psfig}
\usepackage{color}
\usepackage{natbib}
\begin{document}

\title{
Oxygen abundance variations in the system of the two blue compact dwarf
galaxies SBS 0335--052E and SBS 0335--052W\thanks{Based on observations
collected at the European Southern Observatory, Chile, ESO program 
76.B-0739.}
}

\author{P.\ Papaderos \inst{1}
\and Y. I.\ Izotov \inst{2}
\and N. G.\ Guseva \inst{2}
\and T. X.\ Thuan\inst{3}
\and K. J.\ Fricke \inst{1}}
\offprints{P. Papaderos, papade@astro.physik.uni-goettingen.de}
\institute{          Institute for Astrophysics, Friedrich-Hund-Platz 1,
                     37077 G\"ottingen, Germany
\and
                     Main Astronomical Observatory,
                     Ukrainian National Academy of Sciences,
                     Zabolotnoho 27, Kyiv 03680,  Ukraine
\and
                     Astronomy Department, University of Virginia,
                     Charlottesville, VA 22903, USA
}

\date{Received \hskip 2cm; Accepted}

\abstract{We present 3.6m ESO telescope spectroscopic observations 
of the system of the two blue compact dwarf galaxies SBS 0335--052\,W and
SBS 0335--052\,E. 
The oxygen abundance in SBS 0335--052W is 12 + log O/H = 7.13 $\pm$ 0.08, confirming 
that this galaxy is the most metal-deficient emission-line galaxy known. 
We find that the oxygen abundance in SBS 0335--052E varies from region to region 
in the range from 7.20 to 7.31, suggesting the presence of an abundance gradient 
over a spatial scale of $\la$ 1 kpc. 
Signatures of early carbon-type Wolf-Rayet stars are detected in cluster \#3
of SBS 0335--052\,E, corresponding to the emission of three to eighteen
WC4 stars, depending on the adopted luminosity of a single WC4 star in the 
C {\sc iv} $\lambda$4658 emission line.
\keywords{galaxies: fundamental parameters -- galaxies: starburst -- galaxies: abundances}
}
\titlerunning{Oxygen abundance variations in SBS 0335--052E and SBS 0335--052W}
\maketitle

\section{Introduction}

\begin{figure}[t]
\hspace*{1.0cm}\psfig{figure=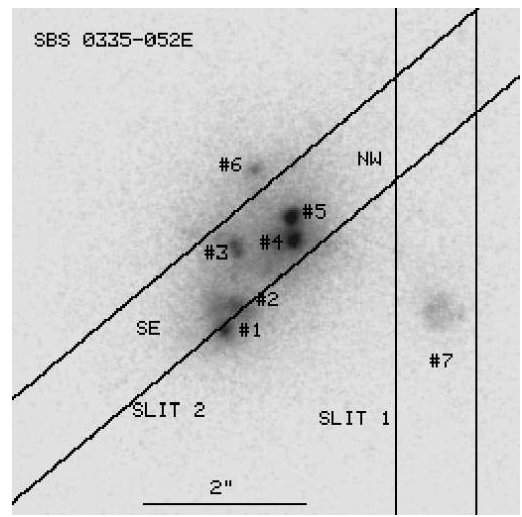,angle=0,width=7.cm,clip=}
\caption{HST/ACS archival UV image of SBS 0335--052E.
The stellar clusters \#1 through \#7 and the SE and NW regions are labeled.}
\label{fig1}
\end{figure}

The system of the two blue compact dwarf (BCD) galaxies SBS 0335--052W and 
SBS 0335--052E is an excellent nearby laboratory for studying star formation 
in low-metallicity environments.
Since its discovery as one of the
most metal-deficient star-forming galaxies known \citep{I90}, 
with oxygen abundance 12 + log O/H $\sim$ 7.30 \citep{M92,I97b,I99,TI05}, 
SBS 0335--052E has often been proposed as a nearby young dwarf 
galaxy \citep{I90,I97b,T97,P98}. This system contains several super-star
clusters \citep{T97,P98}, producing extended regions of ionized gas 
emission \citep{M92,I97b,P98,P04}.
Its companion galaxy SBS 0335--052W, the lowest metallicity emission-line
galaxy known with 12 + log O/H = 7.12 $\pm$ 0.03 \citep{I05a}, is located
at a projected distance of 22 kpc from SBS 0335--052E. It was 
discovered by \citet{P97} as a dwarf emission-line galaxy associated with   
the brightest of the two intensity peaks of 21 cm emission in the 
large (66 by 22 kpc)  H {\sc i} envelope surrounding  
SBS 0335--052E, with which the other H {\sc i} peak is associated  
\citep{P01}.

\begin{figure*}[t]
\hspace*{1.0cm}\psfig{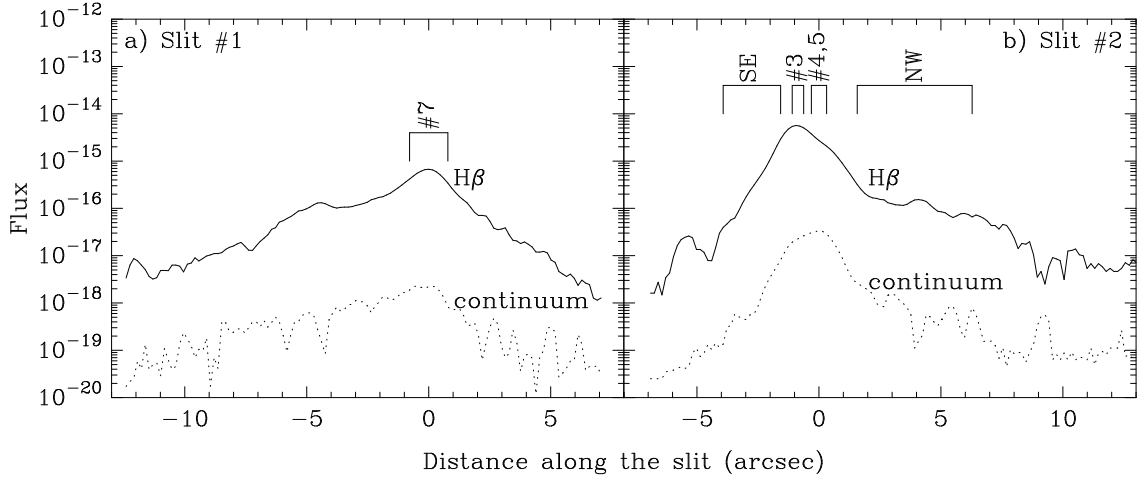}
\caption{Flux distributions in the H$\beta$ emission line (solid line)
and in the continuum (dotted line) along slits 1 (panel a) and 2
(panel b). The H$\beta$ flux is in units of erg s$^{-1}$ cm$^{-2}$ 
and the continuum flux in units of erg s$^{-1}$ cm$^{-2}$ \AA$^{-1}$.
The various regions that have been used to extract one-dimensional spectra are 
labeled.}
\label{fig2}
\end{figure*}

\begin{figure*}[t]
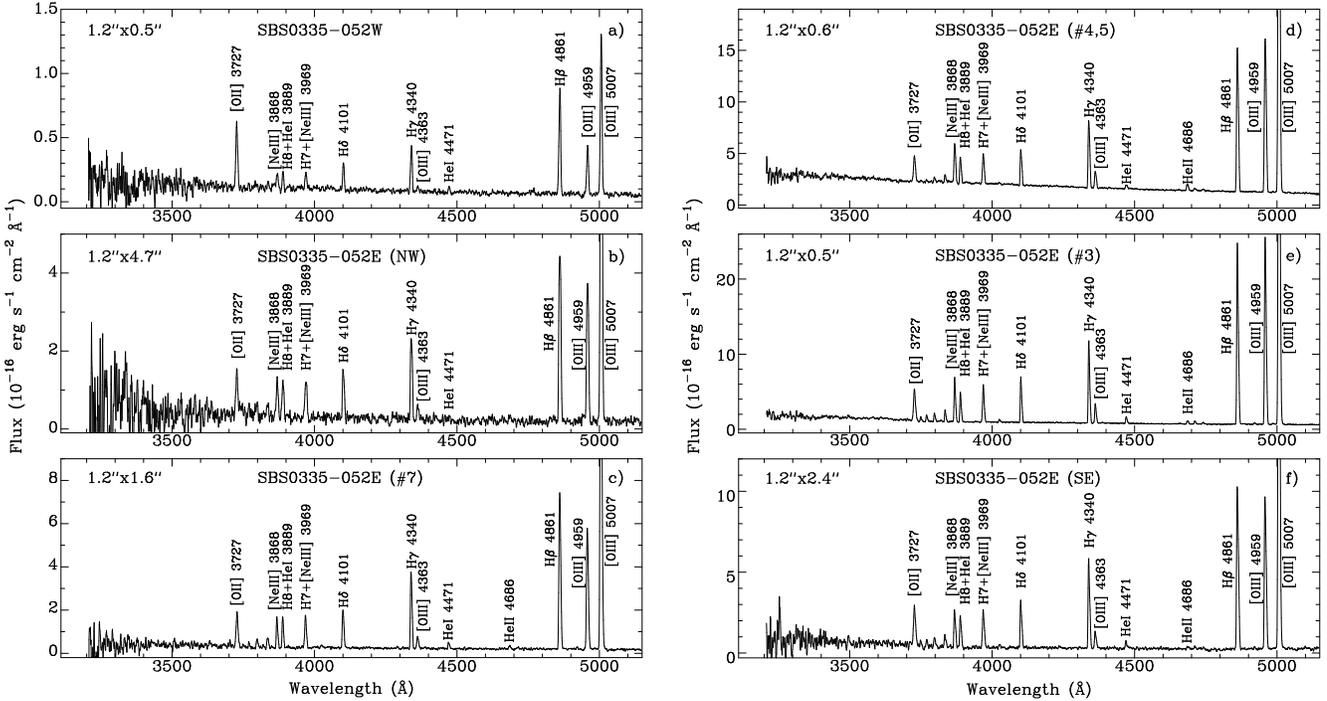

\hspace*{0.0cm}\psfig{figure=5110fig3a.ps,angle=0,width=8.5cm,clip=}
\hspace*{0.4cm}\psfig{figure=5110fig3b.ps,angle=0,width=8.5cm,clip=}
\caption{Spectra of SBS 0335--052W (a) and of different regions in SBS 0335--052E
(b-f).}
\label{fig3}
\end{figure*}

Because of the great interest of the SBS 0335--052 system
to studies of dwarf galaxy formation and evolution, we have carried out
new spectroscopic observations of both BCDs. Our work has been motivated by
the following considerations. 
First, previous spectroscopic studies of SBS 0335--052E have been focused
on the brightest clusters \#1,\#2 and \#4,\#5, indicated on the 
HST/ACS archival UV image\footnote{Based on 
observations made with the NASA/ESA Hubble Space Telescope, obtained from the 
Data Archive at the Space Telescope Science Institute, which is operated by 
the Association of Universities for Research in Astronomy, Inc., under NASA 
contract NAS 5-26555. These observations are associated with programme \# 9470.} (Fig. \ref{fig1}). 
No abundance determination has been done for fainter clusters.
In this paper we derive for the first time oxygen abundances 
in the regions around clusters \#3 and \#7. Second, \citet{I97b}
have found an oxygen abundance gradient along the major axis of
SBS 0335--052E, with the highest abundance associated with the brightest 
cluster \#1. 
We wish to study whether oxygen abundance variations are present in other
regions of SBS 0335--052E as well. 
Third, we wish to check whether SBS 0335--052W
is indeed the lowest-metallicity emission-line galaxy known. 

\begin{table*}
\caption{Extinction-corrected Emission Line Fluxes \label{tab1}}
\begin{tabular}{lcccccc} \hline
  &  & \multicolumn{5}{c}{0335--052E} \\ \cline{3-7}
Line                           &0335--052W     &         NW     &         \#7   &         \#4,5 &         \#3   &         SE     \\ \hline
3727 [O {\sc ii}]              &0.646$\pm$0.026 &0.256$\pm$0.019&0.259$\pm$0.009&0.214$\pm$0.006&0.282$\pm$0.005&0.251$\pm$0.007 \\
3750 H12                       &      ...       &      ...      &      ...      &      ...      &0.027$\pm$0.003&0.040$\pm$0.009 \\
3771 H11                       &      ...       &      ...      &0.054$\pm$0.013&0.041$\pm$0.012&0.039$\pm$0.003&0.048$\pm$0.008 \\
3797 H10                       &      ...       &      ...      &0.078$\pm$0.011&0.059$\pm$0.008&0.054$\pm$0.003&0.065$\pm$0.008 \\
3820 He {\sc i}                &      ...       &      ...      &      ...      &      ...      &0.011$\pm$0.002&      ...       \\
3835 H9                        &      ...       &      ...      &0.089$\pm$0.009&0.070$\pm$0.006&0.071$\pm$0.003&0.076$\pm$0.008 \\
3869 [Ne {\sc iii}]            &0.122$\pm$0.013 &0.216$\pm$0.013&0.173$\pm$0.006&0.272$\pm$0.005&0.271$\pm$0.005&0.202$\pm$0.006 \\
3889 He {\sc i} + H8           &0.188$\pm$0.028 &0.215$\pm$0.019&0.198$\pm$0.008&0.206$\pm$0.006&0.185$\pm$0.004&0.181$\pm$0.008 \\
3968 [Ne {\sc iii}] + H7       &0.208$\pm$0.023 &0.269$\pm$0.020&0.230$\pm$0.010&0.242$\pm$0.006&0.250$\pm$0.005&0.233$\pm$0.008 \\
4026 He {\sc i}                &      ...       &      ...      &      ...      &      ...      &0.017$\pm$0.001&0.023$\pm$0.004 \\
4101 H$\delta$                 &0.260$\pm$0.020 &0.274$\pm$0.017&0.260$\pm$0.008&0.265$\pm$0.006&0.264$\pm$0.004&0.259$\pm$0.007 \\
4340 H$\gamma$                 &0.476$\pm$0.021 &0.475$\pm$0.018&0.476$\pm$0.010&0.475$\pm$0.008&0.473$\pm$0.007&0.478$\pm$0.009 \\
4363 [O {\sc iii}]             &0.047$\pm$0.010 &0.081$\pm$0.008&0.079$\pm$0.004&0.118$\pm$0.003&0.111$\pm$0.002&0.095$\pm$0.004 \\
4471 He {\sc i}                &0.039$\pm$0.010 &     ...       &0.037$\pm$0.003&0.031$\pm$0.002&0.036$\pm$0.001&0.039$\pm$0.003 \\
4658 [Fe {\sc iii}]            &      ...       &      ...      &      ...      &0.004$\pm$0.002&      ...      &      ...       \\
4686 He {\sc ii}               &      ...       &      ...      &      ...      &0.046$\pm$0.002&0.021$\pm$0.001&0.021$\pm$0.004 \\
4711 [Ar {\sc iv}] + He {\sc i}&      ...       &     ...       &0.018$\pm$0.004&0.017$\pm$0.002&0.016$\pm$0.001&0.019$\pm$0.003 \\
4740 [Ar {\sc iv}]             &      ...       &     ...       &      ...      &0.012$\pm$0.002&0.011$\pm$0.001&0.015$\pm$0.003 \\
4861 H$\beta$                  &1.000$\pm$0.025 &1.000$\pm$0.023&1.000$\pm$0.017&1.000$\pm$0.016&1.000$\pm$0.015&1.000$\pm$0.016 \\
4921 He {\sc i}                &      ...       &     ...       &0.012$\pm$0.002&0.010$\pm$0.005&0.007$\pm$0.001&0.012$\pm$0.003 \\
4959 [O {\sc iii}]             &0.465$\pm$0.016 &0.801$\pm$0.018&0.790$\pm$0.013&1.084$\pm$0.017&1.075$\pm$0.016&0.939$\pm$0.015 \\
4988 [Fe {\sc iii}]            &      ...       &      ...      &      ...      &0.006$\pm$0.002&0.003$\pm$0.001&      ...       \\
5007 [O {\sc iii}]             &1.435$\pm$0.033 &2.404$\pm$0.046&2.410$\pm$0.037&3.187$\pm$0.048&3.245$\pm$0.047&2.891$\pm$0.044 \\ \\
$C$(H$\beta$)                  &      0.025     &     0.160     &     0.140     &    0.260      &     0.430     &    0.000       \\
EW(H$\beta$)$^a$                   &      109       &    209        &    299        &      89       &     246       &     382        \\
$F$(H$\beta$)$^b$              &      6.3       &    39.6       &    56.7       &   113.7       &   310.2       &    82.4        \\
EW(abs)$^a$                        &      3.4       &     1.1       &      5.0      &     1.2       &     0.1       &     2.6        \\
\hline
\end{tabular}

$^a$ in \AA.

$^b$ in units 10$^{-16}$ erg s$^{-1}$ cm$^{-2}$.

\end{table*}

\begin{table*}[t]
\caption{Ionic and Total Element Abundances \label{tab2}}
\begin{tabular}{lcccccc} \hline
  &  & \multicolumn{5}{c}{0335--052E} \\ \cline{3-7}
Property                          &0335--052W     &         NW     &         \#7   &         \#4,5 &         \#3   &         SE     \\ \hline
$T_e$(O {\sc iii}), K           &19710$\pm$2460 &19950$\pm$1210  &19740$\pm$550  &21110$\pm$390  &20220$\pm$260  &19780$\pm$490   \\
$T_e$(O {\sc ii}), K            &15590$\pm$1800 &15610$\pm$\,~870&15590$\pm$400  &15630$\pm$260  &15620$\pm$190  &15590$\pm$360   \\ \\
O$^+$/H$^+$, ($\times$10$^5$)     &0.521$\pm$0.153&0.206$\pm$0.033 &0.209$\pm$0.015&0.172$\pm$0.009&0.225$\pm$0.008&0.203$\pm$0.013 \\
O$^{2+}$/H$^+$, ($\times$10$^5$)  &0.839$\pm$0.239&1.381$\pm$0.190 &1.409$\pm$0.089&1.642$\pm$0.069&1.811$\pm$0.057&1.680$\pm$0.097 \\
O$^{3+}$/H$^+$, ($\times$10$^7$)  &      ...      &      ...       &1.995$\pm$0.664&5.015$\pm$0.102&2.193$\pm$0.274&1.910$\pm$0.794 \\
O/H, ($\times$10$^5$)             &1.360$\pm$0.283&1.586$\pm$0.193 &1.637$\pm$0.091&1.863$\pm$0.071&2.058$\pm$0.057&1.901$\pm$0.098 \\
12+log O/H                        &7.13$\pm$0.08  &7.20$\pm$0.05   &7.21$\pm$0.02  &7.27$\pm$0.02  &7.31$\pm$0.01  &7.28$\pm$0.02   \\ \\
Ne$^{2+}$/H$^+$, ($\times$10$^6$) &1.591$\pm$0.455&2.756$\pm$0.389 &2.257$\pm$0.152&3.045$\pm$0.128&3.343$\pm$0.105&2.621$\pm$0.162 \\
$ICF$(Ne)                         &      1.16     &      1.06      &     1.06      &     1.05      &     1.05      &     1.05       \\
Ne/H, ($\times$10$^6$)            &1.851$\pm$0.738&2.907$\pm$0.447 &2.391$\pm$0.176&3.199$\pm$0.145&3.513$\pm$0.120&2.751$\pm$0.183 \\
log Ne/O                          &--0.87$\pm$0.20&--0.74$\pm$0.09 &--0.84$\pm$0.04&--0.77$\pm$0.03&--0.77$\pm$0.02&--0.84$\pm$0.04 \\ 
\hline
\end{tabular}
\end{table*}

\section{Observations and Data Reduction}

New spectra of SBS 0335--052W and SBS 0335--052E were 
obtained on 6 and 8 October, 2005 with the EFOSC2 
(ESO Faint Object Spectrograph and Camera) mounted at the 3.6m ESO 
telescope at La Silla. The observing conditions were photometric 
during these two nights.
All observations were performed with the same instrumental setup. 
We used the $\#07$ ($\lambda$$\lambda$3200--5200) grism 
with the 600 gr/mm grating and a long slit with a width of 1\farcs2. 
This resulted in a spatial scale along the slit of 0\farcs157 pixel$^{-1}$, and a
spectral resolution of $\sim$6.2~\AA\ (FWHM).

Both galaxies were observed at low airmass $\la$ 1.2, so no correction 
for atmospheric refraction has been applied. The seeing was 0\farcs7 -- 0\farcs8.
The data were reduced with the IRAF\footnote{IRAF is 
the Image Reduction and Analysis Facility distributed by the 
National Optical Astronomy Observatory, which is operated by the 
Association of Universities for Research in Astronomy (AURA) under 
cooperative agreement with the National Science Foundation (NSF).}
software package.

The spectrum of the brightest region of SBS 0335--052W was obtained with 
the slit oriented along the parallactic angle and 
with an exposure time of 3600 s.
SBS 0335--052E was observed in the two slit positions shown in Fig. \ref{fig1}.
Slit 1 was
centered on the isolated cluster \#7 and oriented along the parallactic angle. 
Slit 2 had a position 
angle of --50$^\circ$ and was centered on clusters \#4 and \#5.
The total exposure times for slits 1 and 2 were respectively 
2400 and 2700 sec.
We note that the morphology and the location of the clusters 
are similar in the UV (Fig. \ref{fig1}) and optical wavelengths \citep[cf. ][]{T97,P98}, so that  
the placement of the slits during the spectroscopic observations, 
based on short acquisition exposures in $V$, 
is correctly represented in Fig. \ref{fig1}.

We show in Fig. \ref{fig2} the distribution of the H$\beta$ emission line
flux (solid line) and that of the continuum flux near H$\beta$ (dotted line),
together with the regions used to extract one-dimensional spectra. 
These are shown in panels 
b) through f) in Fig. \ref{fig3}.
Examination of panel b in Fig. \ref{fig2} reveals that near  
the axis origin, the flux distribution of the H$\beta$ line
is slightly offset (by $\sim$ 1\arcsec) relative to that of the continuum, 
with the first distribution peaking on cluster \#3 and the second one 
peaking on clusters \#4 and \#5.  
This shift is also reflected in the equivalent width of the 
H$\beta$ line of clusters \#4 and \#5 which, at 89 \AA, is almost 
three times lower than the one of cluster \#3 (cf. Table \ref{tab1}).
The fact that the maximum of the optical continuum emission coincides 
with the position of clusters \#4 and \#5 ensures that 
the visually chosen center of slit 2 
was not offset due to strong off-center nebular line emission.

Despite the small angular separation between the various clusters 
in SBS 0335--052E, contamination of cluster \#3 by
the fainter emission from the H {\sc ii} region around  
clusters \#4 and \#5 and from the northeastern periphery of clusters \#1 and \#2 is 
relatively low, as can be seen from Figs. \ref{fig1} and \ref{fig2}b.
However, it is likely that contamination of clusters \#4 and \#5
by the emission from the H {\sc ii} region associated with cluster \#3 
is more important, making the abundance determination of the 
H {\sc ii} region associated with clusters \#4 and \#5 less reliable.  

The flux-calibrated spectra of the brightest region of SBS 0335--052W and of
different regions in SBS 0335--052E are shown in Fig. \ref{fig3}.
The aperture within which each spectrum was extracted is given 
in the upper-left part of each panel.
The spectra were corrected for interstellar extinction using the 
reddening curve of \citet{W58} and for redshift. 

Emission-line fluxes were measured using Gaussian profile fitting. 
The errors of the line fluxes were calculated from the photon statistics
in the non-flux-calibrated spectra. They have been propagated in the 
calculations of the elemental abundance errors.
The extinction-corrected emission line fluxes $I$($\lambda$) relative to the 
H$\beta$ fluxes, the extinction coefficients $C$(H$\beta$), the 
equivalent widths EW(H$\beta$),
the observed H$\beta$ fluxes $F$(H$\beta$) and the 
equivalent widths EW(abs) of the hydrogen absorption lines 
are listed in Table \ref{tab1}. 

\section{Results}

\subsection{Physical Conditions and Element Abundances}

The electron temperature $T_{\rm e}$, the 
ionic and total heavy element abundances were derived 
following \citet{I05b}. In particular for the 
O$^{2+}$ and Ne$^{2+}$ ions, we adopt
the temperature $T_e$(O {\sc iii}) derived from the 
[O {\sc iii}] $\lambda$4363/($\lambda$4959 + $\lambda$5007)
emission-line ratio. The O$^+$ abundances were derived with the temperature
$T_e$(O {\sc ii}). The latter was obtained from the relation between $T_e$(O {\sc iii})
and $T_e$(O {\sc ii}) derived by \citet{I05b}.
Observations with the 3.6m telescope were obtained in the blue region,
so the [S {\sc ii}] $\lambda$6717, 6731 emission lines 
usually used for the determination
of the electron number density were out of the observed spectral range. 
Therefore for abundance determinations, we have adopted 
$N_e$ = 100 cm$^{-3}$. The precise value of the electron number density 
makes little difference in the derived abundances
since in the low-density limit which holds for the H {\sc ii} regions
considered here, the element abundances do not depend sensitively 
on $N_e$.
The electron temperatures $T_{\rm e}$(O {\sc iii}) 
and $T_{\rm e}$(O {\sc ii}) respectively 
for the high- and low-ionization zones in the H {\sc ii} regions,
the ionization correction factors ($ICF$s), 
the ionic and total oxygen and neon abundances 
are given in Table \ref{tab2}.

The oxygen abundance in SBS 0335--052W
is 12 + log O/H = 7.13 $\pm$ 0.08. Thus, despite the 
faintness of this galaxy and the 
large errors in its abundance determination, 
the derived oxygen abundance is consistent with the value 
12 + log O/H = 7.12 $\pm$ 0.03 found by \citet{I05a} from higher 
signal-to-noise ratio observations.
The highest oxygen abundance in SBS 0335--052E, 12 + log O/H = 7.31 $\pm$ 0.01,
is determined at the H {\sc ii} region around cluster \#3. This region is also 
characterized by the highest reddening, $C$(H$\beta$) = 0.43.
The derived oxygen abundance is the same as the value 
12 + log O/H = 7.31 $\pm$ 0.01
found by \citet{TI05} for the brightest part (clusters \#1 and \#2) 
of SBS 0335--052E.
The oxygen abundances in other regions of SBS 0335--052E are lower,
with the lowest values occurring in the NW region (7.20 $\pm$ 0.05) and in the H {\sc ii}
region around cluster \#7 (7.21 $\pm$ 0.02). 
These low values are comparable to the oxygen
abundances of 7.17 $\pm$ 0.01 and 7.22 $\pm$ 0.02 determined
respectively by \citet{TI05} for the NW and SE components of the 
BCD I\ Zw\ 18. 
Thus, it is clear that the oxygen abundance in SBS 0335--052E 
shows variations of up to 0.1 dex on spatial scales of $\la$ 1 kpc
(1\arcsec\ corresponds to $\sim$ 260 pc at the adopted distance 
of $D$ = 54.3 Mpc to SBS 0335--052E).

Our results suggest a trend for the oxygen abundance in SBS 0335-052E
to decrease from the brightest part of the galaxy to its outer fainter 
parts, in agreement with the finding of \citet{I97b}. 
\citet{T97} and \citet{P98} have found 
a systematic increase in the $V-I$ colours of the clusters away from the 
optically brightest ones (clusters \#1 and \#2). 
This trend is probably due to the combined effect of differential 
extinction by dust 
and the colour evolution of stellar clusters formed sequentially  
in a propagating star 
formation process from the NW region of the galaxy to its SE region. 
Clusters \#1, \#2 and \#3 are 
the youngest ones, while clusters \#4 and \#5 are older, as evidenced by 
their lower H$\beta$ equivalent width.
In that scenario, the larger abundances of clusters \#1, \#2 and \#3 may 
be interpreted as an effect of self-enrichment \citep{KS86} in these young clusters.

The Ne/O abundance ratios in the two BCDs are consistent within the errors with the ratios
obtained for other metal-deficient galaxies \citep{IT99}.

\subsection{Wolf-Rayet Stars}

\begin{figure}[t]
\hspace*{0.5cm}\psfig{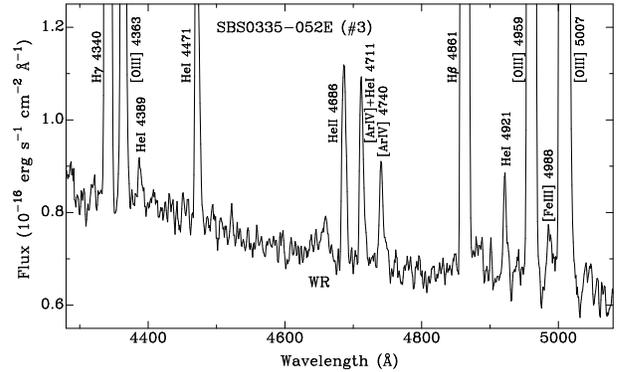}
\caption{Spectrum of cluster \#3 in SBS 0335--052E with the 
Wolf-Rayet (WR) emission feature indicated.}
\label{fig4}
\end{figure}

In the spectrum of cluster \#3 (Fig. \ref{fig4}), we detect a broad emission
feature at $\lambda$ $\sim$ 4658 \AA\ (labeled ``WR''). 
This feature is coincident with the [Fe {\sc iii}]$\lambda$4658 nebular 
emission line but its width is much larger than those of other 
nebular emission 
lines in the spectrum of cluster \#3. 
We interpret therefore the broad feature as due mainly 
to the stellar Wolf-Rayet (WR) emission line 
C {\sc iv} $\lambda$4658, a signature of WC4 stars. 
This would make SBS 0335--052E the second most metal-deficient galaxy, after
I Zw 18 \citep{I97a,L97}, in which a WR stellar population has been detected.
The total $\lambda$4658 flux, corrected for the interstellar 
extinction $C$(H$\beta$) = 0.43, is equal
to 4.8$\times$10$^{-16}$ erg s$^{-1}$ cm$^{-2}$. 
To correct for the contribution of the 
[Fe {\sc iii}] $\lambda$4658 emission line, we use the flux of 
another iron emission line seen 
in the spectrum of cluster \#3, [Fe {\sc iii}] $\lambda$4988 
(Table \ref{tab1}, Fig. \ref{fig4}).  Adopting 
the mean [Fe {\sc iii}] $\lambda$4988/$\lambda$4658 flux ratio 
to be 1.5, as obtained from the BCD sample of \citet{TI05},
the [Fe {\sc iii}]$\lambda$4658 flux is about 
1.7$\times$10$^{-16}$ erg s$^{-1}$ cm$^{-2}$. Correcting for it gives 
$I$(C {\sc iv}$\lambda$4658) = 3.1$\times$10$^{-16}$ erg s$^{-1}$ cm$^{-2}$.
This 
corresponds to a luminosity $L$(C {\sc iv}) = 
1.4$\times$10$^{38}$ erg s$^{-1}$ 
or to an equivalent number of WC4 stars $N$(WC4) = 3, if we adopt   
5$\times$10$^{36}$ erg s$^{-1}$ \citep{SV98,G00} for the luminosity of 
one WC4 star.
The number of O7{\sc v} stars derived from the H$\beta$ luminosity,
following \citet{G00}, is
$N$(O7{\sc v}) $\sim$ 800, giving a WC4-to-O7{\sc v} number ratio of 
$\sim$ 0.003.
This number ratio is consistent with the one predicted by stellar population
synthesis models at the metallicity of SBS 0335--052E \citep{SV98}.
The non-detection of WR features in previous spectroscopic studies of SBS 0335-052E which focused on clusters \#1,2 and \#4,5, 
can probably be understood as a result of the very short
duration of the WR stage ($<$1 Myr) in low-metallicity stellar clusters, 
as only the most massive  
short-lived stars go through this stage. 
We are therefore just lucky to catch stars going through 
the WR stage in cluster \#3.
We note that a similar situation holds in  
I\ Zw\ 18: \citet{Brown02} have detected WR stars only
in two very compact clusters associated with the 
NW star-forming component of the galaxy, but WR features are not seen  
in its SE component.

Recently \citet{CH05} have shown that the line luminosities of extremely 
low-metallicity WC4 stars, such as those observed in cluster \#3 
and in I Zw 18, are 3 -- 6 times lower than those predicted 
by \citet{SV98}. This would result in an increase of $N$(WC4) to 9 -- 18.
A further 
search for Wolf-Rayet stellar populations in cluster \#3 
as well as in other regions of SBS 0335-052E would be of great interest.
To confirm the detection of WC4 stars and to better determine their number,
better 
observations of the broad C {\sc iv} $\lambda$5808 emission line are necessary.
It is also important to search for spectroscopic features of 
late nitrogen WR (WNL) stars which, from stellar evolutionary considerations, 
are likely present in cluster \#3.
The weakness of the N {\sc iii} $\lambda$4640 and 
He {\sc ii} $\lambda$4686 emission lines and the 
absence of appreciably broad emission in those lines 
(Fig. \ref{fig4}) precludes the determination of $N$(WNL) 
from the present spectra. 
Nevertheless, we can conclude from our data that Wolf-Rayet stars 
are present in cluster \#3 of SBS 0335-052E, and that their number 
is likely greater than $\sim$10.

\section{Summary}

We present 3.6m ESO telescope spectroscopic observations of the system of the 
two extremely metal-deficient blue compact dwarf (BCD)
galaxies SBS 0335--052W and SBS 0335--052E.
The oxygen abundance in SBS 0335--052W is 12 +log O/H = 7.13 $\pm$ 0.08,
confirming the earlier finding by \citet{I05a} that this BCD is the most
metal-deficient emission-line galaxy known. 
As for SBS 0335--052E, its oxygen abundance varies
from region to region between 7.20 and 7.31, suggesting a metallicity gradient
over a spatial scale of $\la$ 1 kpc. We thus confirm and extend the finding
by \citet{I97b} of an abundance gradient in SBS 0335--052E.
In cluster \#3 of SBS 0335--052E, we find evidence for the presence of 
early carbon-type Wolf-Rayet (WR) stars, making this galaxy the second 
most-metal deficient galaxy with a detected WR stellar population, after 
I Zw 18.

\begin{acknowledgements}
P.P. would like to thank Gaspare Lo Curto, Lorenzo Monaco, Carlos La Fuente, 
Eduardo Matamoros and the whole ESO staff at the La Silla Observatory for 
their support.
Y. I. I. and N. G. G. thank the hospitality of the Institute for Astrophysics
(G\"ottingen), the support of the DFG grant No. 436 UKR 17/25/05 and of the 
grant No 02.07.00132 from the  Ukrainian Fund of Fundamental Investigations.
Y. I. I. and T. X. T. acknowledge the partial financial support of NSF
grant AST 02-05785. 
T. X. T. thanks the hospitality of the Institut d'Astrophysique in Paris 
and of the Service d'Astrophysique at Saclay during his sabbatical leave.
He is grateful for a Sesquicentennial Fellowship from the University 
of Virginia. 
The research described in this publication was made
possible in part by Award No. UP1-2551-KV-03 of the US Civilian Research
\& Development Foundation for the Independent States of the Former
Soviet Union (CRDF).
\end{acknowledgements}

\end{document}